\documentclass[twoside,superscriptaddress,twocolumn,floatfix,a4paper,aps]{revtex4-1}

\usepackage{color}
\usepackage{graphicx}
\usepackage[utf8x]{inputenc}
\usepackage[T1]{fontenc}
\usepackage{siunitx}
\usepackage{amstext}
\usepackage{textgreek}
\usepackage{hyperref}
\usepackage{epstopdf}
\usepackage{amsfonts}
\usepackage{amsmath}
\usepackage{multirow}
\usepackage{dsfont}

\usepackage{marvosym}
\usepackage[rgb]{xcolor}

\makeatletter

\newcommand{\Rnum}[1]{\expandafter\@slowromancap\romannumeral #1@}

\begin{document}

\title{Experimental measurement of nonlinear entanglement witness\\ by hyper-entangling two-qubit states}

\author{Vojtěch Trávníček}
\email{vojtech.travnicek@upol.cz}
\affiliation{RCPTM, Joint Laboratory of Optics of Palacký University and Institute of Physics of Czech Academy of Sciences, 17. listopadu 12, 771 46 Olomouc, Czech Republic}

\author{Karol Bartkiewicz} \email{bark@amu.edu.pl}
\affiliation{RCPTM, Joint Laboratory of Optics of Palacký University and Institute of Physics of Czech Academy of Sciences, 17. listopadu 12, 771 46 Olomouc, Czech Republic}
\affiliation{Faculty of Physics, Adam Mickiewicz University,
PL-61-614 Pozna\'n, Poland}

\author{Antonín Černoch} \email{acernoch@fzu.cz}
\affiliation{Institute of Physics of the Czech Academy of Sciences, Joint Laboratory of Optics of PU and IP AS CR, 17. listupadu 50A, 772 07 Olomouc, Czech Republic}
   
\author{Karel Lemr}
\email{k.lemr@upol.cz}
\affiliation{RCPTM, Joint Laboratory of Optics of Palacký University and Institute of Physics of Czech Academy of Sciences, 17. listopadu 12, 771 46 Olomouc, Czech Republic}   

\begin{abstract}
We demonstrate that  non--linear entanglement witnesses can be made particularly useful for entanglement detection in hyper--entangled or multilevel states. We test this idea experimentally on the platform of linear optics using a hyper--entangled state of two photons. Instead of several simultaneous copies of two-photon entangled states, one can directly measure the witness on single copy of a hyper--entangled state. Our results indicate that hyper--entanglement can be used for  quick entanglement detection  and it provides a practical testbed for experiments with non--linear entanglement witnesses.
\end{abstract}

\date{\today}

\maketitle
\section{Introduction}

Quantum entanglement is one of the most peculiar phenomenon in quantum physics \cite{Mintert,Horodecki1}. It is also the cornerstone of a number of quantum technologies \cite{Briegel,Ekert,Hillery}. More specifically, it plays a crucial role in quantum communications and quantum computing \cite{Riedmatten, Barenco}. Therefore it is no surprise that quantum entanglement has been subject of an intense investigation \cite{Horodecki1,Plenio}.

There are two commonly used methods for entanglement characterization. The first method is based on quantum state tomography and density matrix reconstruction \cite{Salles, Miranowicz, Bartkiewicz, Jezek}. The advantage of this strategy is that one does not have to possess any \emph{a priory} information about the input state. On the other hand, executing a full state tomography is a time demanding process especially for multilevel quantum states. The number of required measurements grows exponentially with the dimension of investigated state which makes both the necessary measurement and the related data processing very time consuming~\cite{Hou}.

The second method is based on the so-called entanglement witnesses (EW). Measuring simple linear EW requires performing a set of suitable local measurements which are direct products of projections applied on each subsystem separately. These projections are chosen based on some \emph{a priory} information about the investigated state. Correlations of these local measurements across involved parties then reveal the entanglement \cite{Clauser}. 

\begin{figure}[!h!]
                \begin{center}
                \includegraphics[scale=1]{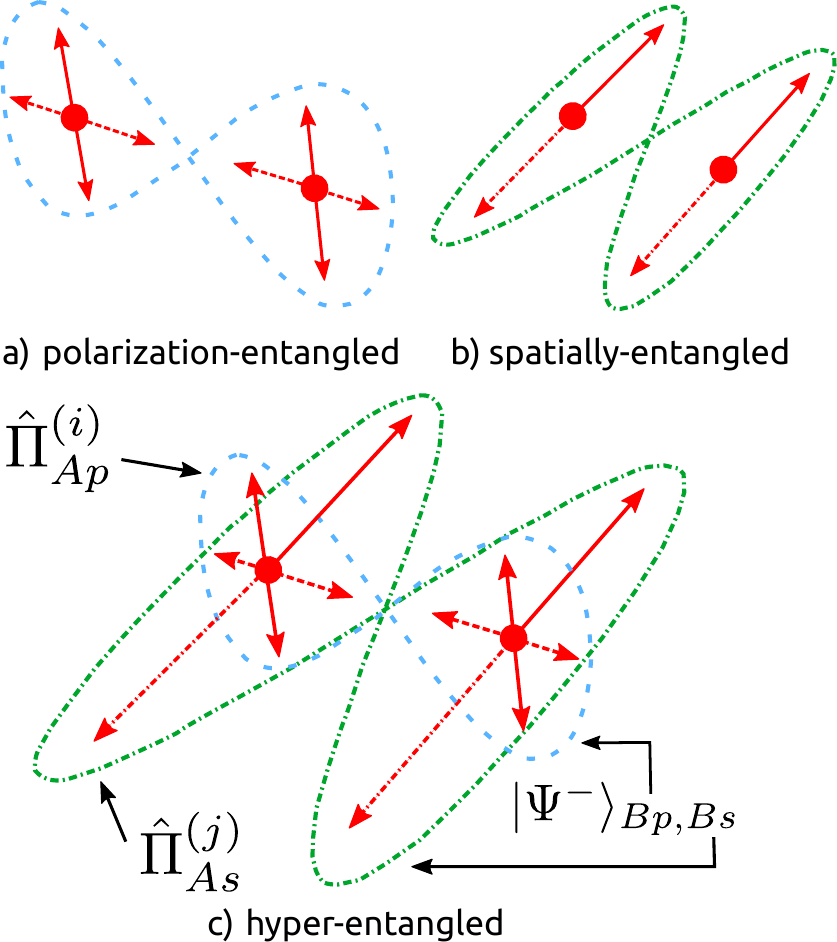}
                \caption{\label{scheme} (color online) Conceptual scheme for measuring the collectibility of two-qubit states by hyper--entangling. The initial state $\hat{\rho}^{(p)} = \hat{\rho}^{(s)}$ is encoded both into polarization (p) and spatial (s) modes of photons A and B. Photon A is then subjected to projections $\hat{\Pi}_{Ap}, \hat{\Pi}_{As}$ that are defined for polarization and spatial mode, respectively. Photon B is projected onto a singlet state $|\psi^{-}\rangle_{Bp,Bs} = \tfrac{1}{\sqrt{2}}(|01\rangle-|10\rangle)_{Bp,Bs}$, where the two degrees of freedom are addressed holistically at the same time.}       
                \end{center}
\end{figure}

The second class of EWs encompasses the nonlinear (collective) entanglement witnesses \cite{BartkiewiczPRA17a,BartkiewiczPRA18,Horodecki2} which remove the need for \emph{a priory} information about a given state, but require simultaneous measurements on at least two copies of the state. This idea has been experimentally demonstrated in a seminal paper by Bovio \emph{et al.} \cite{Bovino}. In fact, a number of nonlinear EWs   \cite{13,14,15,16,17,18,19,20,21,22,23,24,25,26,27} have been devised for various classes of quantum states. Moreover,  universal experimental optical approaches to measuring or detecting the entanglement of an arbitrary two-qubit state have been reported~\cite{BartkiewiczPRA15a,BartkiewiczPRA15b,BartkiewiczPRA17b}. These universal approaches require using up to four copies of the state.

In this paper we promote the benefits of entanglement detection by means of a nonlinear EW on hyper--entangled states (HESs). A HES is a quantum state entangled in more then one degree of freedom (DOF) and can be written in the form 
$$\hat{\rho}_{\mathrm{HES}}=\hat{\rho}^{(p)} \otimes \hat{\rho}^{(s)},$$
where subscripts $p$ and $s$ stand for two independent DOFs. These states are an invaluable resource for quantum information protocols. They can be used for increasing channel capacity \cite{Barreiro}, efficient quantum key distribution \cite{Perumangatt}, two-qubit teleportation \cite{Perumangatt} or serve as a powerful  safeguard against eavesdropping \cite{Jiarui}. Therefore, their quick detection and diagnosis is of paramount importance for practical implementation of the mentioned protocols. By hyper-entangling a standard two-photon polarization state instead of preparing multiple copies of  a polarization-entangled pair of photons, we  measure a nonlinear EW on a single pair of hyper--entangled photons. For the purposes of demonstrating this concept, we have selected one of the less complex nonlinear EWs known as the collectibility \cite{21,23}.

The original technique proposed by the authors of Ref.~\cite{21,23} allows to detect two-qubit entanglement in a large number of entangled states without the need for any \emph{a priory} information about the investigated state. The collectibility measurement of two-qubit systems entangled in one DOF requires to perform the collective measurements on two identical copies of the investigated state \cite{col1d}. On the other hand collectibility can be measured directly on a single multilevel state where a two-qubit state is copied across two DOFs forming a HES. The collective measurements of HES consists of local and non--local projections. Local and non--local in our concept stand for projections implemented separately respectively across the two DOFs (see the conceptual scheme in Fig.~\ref{scheme}). We test this method on three characteristic two-qubit quantum states encoded twice in two separate DOFs of a pair of photons. These states are a Bell state, a pure separable state and a maximally mixed state.

We demonstrate that the HES method developed in this paper is experimentally feasible on the platform of linear optics. Also because the method requires using only a single pair of photons to encode two copies of the investigated state, it can be applied for fast and easy to implement diagnostics of a given HES. With linear optics, simultaneous measurement of multiple copies of a pair of photons is quite time-consuming as the generation rate decreases exponentially with the number of copies. Moreover, the experimental friendliness of our method makes it a suitable testbed for \textbf{further} development of other more powerful nonlinear EWs or it can be easily adapted to measure a class of two-copy based EWs studied recently in Ref.~\cite{BartkiewiczPRA17a}.

\section{Theoretical framework}
\label{theory}
A hyper--entangled state (HES) is, e.g., a system composed of two quantum particles (subsystems $A$ and $B$) each encoding two qubits, one qubit per a degree of freedom (DOF). A pair of photons with polarization (p) and spatial (s) DOFs  is an example of such system. For the purpose our analysis we consider only identical states of encoded into both DOFs, i.e., $\hat{\rho}^{(p)} = \hat{\rho}^{(s)}$, where $\hat{\rho}^{(x)}=\hat\rho_{Ax,Bx}$ for $x =p,s$.

The collectibility of such system is defined in terms of four different local projections implemented simultaneously with one nonlocal projection \cite{col1d}. One subsystem ($A$) of the HES is measured with separable projections, the other ($B$) with an entangled projection. The amount of entanglement, in terms of collective nonlinear entanglement witness $W(\hat{\rho})$, is then derived from correlations between coincidence rates observed for these projections. The correlations between coincidence rates of individual projections were labeled $p_{ij}$ can be expressed in terms of joint projection probabilities
\begin{equation}
\begin{split}
p_{ij} & = \mathrm{tr}\left[  \hat{\Pi}^{(i)}_{Ap}\otimes \hat{\Pi}^{(j)}_{As}\otimes\left(|\psi^{-}\rangle\langle \psi^{-}|\right)_{Bp,Bs}\,\hat\rho_{\mathrm{HES}}\right].
\end{split}
\label{projections}
\end{equation}
Indices $Xp$ and $Xs$ mark the relevant DOFs for subsystem $X=A, B$, whereas indices $i,j = 0,1,+$ represent projections on qubit states $|0\rangle$, $|1\rangle$, and $|+\rangle=(|0\rangle+|1\rangle)/\sqrt{2}$ expressed in the computational basis. The projection $(|\psi^{-}\rangle\langle \psi^{-}|)_{Bp,Bs}$ stands for projecting the subsystem $B$ onto a singlet state across the DOFs $p$ and $s$, where
\begin{equation}
|\psi^{-}\rangle_{Bp,Bs} = \tfrac{1}{\sqrt{2}}\left(|01\rangle - |10\rangle\right)_{Bp,Bs}.
\label{eq:singlet}
\end{equation}
Using the notation from Ref.~\cite{col1d} and Eqs.~(\ref{projections}), the collective nonlinear entanglement witness can be formulated as
\begin{equation}
\begin{split}
W(\hat{\rho}) = & \frac{1}{2}[\eta + P^{2}(1-p_{00}) + (1-P)^{2}(1-p_{11})\\
                & + 2P(1-P)(1-p_{01})-1],
\end{split}
\end{equation}
where 
\begin{equation}
\eta = 16P(1-P)\sqrt{p_{00}p_{11}} + 4p_{++},
\end{equation}
and $P$ is the probability of observing subsystem $A$ in state $|00\rangle_{Ap,As}$ independently of the state of subsystem $B$.

\section{Experimental implementation}

\begin{figure}[!htb]
                \begin{center}
                \includegraphics[scale=1]{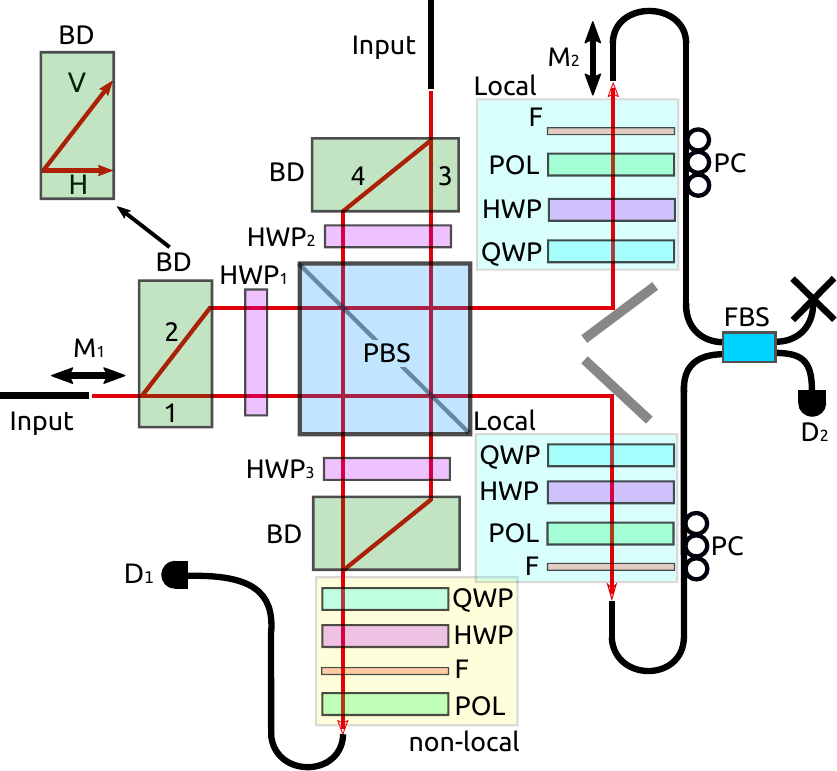}
                \caption{\label{setup} (color online) Experimental setup for measuring collectibility of photonic two-qubit states by hyper--entangling the input photons. M: motorized translation, BD: beam displacer, HWP: half-wave plate, PBS: polarization beam splitter, QWP: quarter-wave plate, POL: polarizer, F: \SI{10}{\nano\meter} interference filter, PC: polarization controller, FBS: fiber beam splitter, D: single-photon detector. Spatial modes are labeled by numbers 1--4.}   
                \end{center}
\end{figure}

The experimental implementation was realized on a platform of linear optics with a hyper-entangled pair of photons encoding the same two-qubit state in both polarization and which-path DOFs. A horizontally polarized photon (i.e., subsystem $X=A,B$) encodes the logical state $|0\rangle_{Xp}$, a vertically-polarized photon state $|1\rangle_{Xp}$. Similarly, its spatial modes $1$ and $3$ encode logical state $|0\rangle_{As}$ and $|0\rangle_{Bs}$, modes $2$ and $4$ logical state $|1\rangle_{As}$ and $|1\rangle_{Bs}$ (see scheme in Fig.~\ref{setup}). Thus,  the relation between the states encoded in the computational basis and in the states used in the experiment can be expressed as
\begin{equation}
\begin{split}
|mn\rangle_{Ap,As} &\equiv  |\delta_{m,1}V+\delta_{m,0}H\rangle_{1+n}\\
|kl\rangle_{Bp,Bs} & \equiv  |\delta_{k,1}V+\delta_{k,0}H\rangle_{3+l},
\end{split}
\label{eq:encode}
\end{equation}
where $\delta$ stands for the Kronecker's delta, indices $m,k=0,1$ mark the single-photon polarization states, and indices $n,l=0,1$ the spatial modes $1+k$ and $3+l$.

The experimental setup consists of a two--photon source powered by pulsed Paladine (Coherent) laser at $\lambda$ = \SI{355}{\nano\meter} with \SI{300}{\milli\watt} of mean optical power and repetition rate of \SI{120}{\mega\hertz}.

Polarization--entangled photon pairs at $\lambda$ = \SI{710}{\nano\meter} are generated in non-collinear type-\Rnum{1} spontaneous parametric down-conversion (SPDC) process in a BBO ($\beta$-BaB$_{2}$O$_{4}$) crystal cascade (known as the Kwiat source \cite{37}). This type of light source if pumped by a generally polarized pumping beam generates pairs of horizontally and vertically polarized photons. Generation rates of these photons as well as their mutual phase shift can be tuned by adjusting the polarization of pumping beam. This way we can prepare states with various amount of entanglement. Each photon from the generated pair is coupled into a single--mode optical fiber and brought to one input port of the experimental setup depicted in Fig.~\ref{setup}.

The beam displacers transform polarization entanglement into spatial entanglement and the two photons can interact on a polarizing beam splitter (PBS) where they get entangled in polarization again. Thus, a given HES is prepared.

The HES is consequently subjected to separable and entangled polarization projective measurements. The photon leading to the detector $D_{2}$ is subjected to separable projections $\hat{\Pi}^{(i)}_{p}\otimes \hat{\Pi}^{(j)}_{s}$ for $i,j=0,1$ as described in Sec.~\ref{theory}. For example projection $\hat{\Pi}^{(0)}_{Ap}\otimes \hat{\Pi}^{(0)}_{As}$ corresponds to a projection onto a state $|00\rangle_{Ap,As}$ or, by using Eq.~(\ref{eq:encode}), $|H\rangle_{1}$. The photon impinging on the detector $D_{1}$ is projected on state $|\psi^{-}\rangle_{Bp,Bs}$  or equivalently [see Eqs.~(\ref{eq:singlet}) and (\ref{eq:encode})] state $\tfrac{1}{\sqrt{2}}(|H\rangle_4-|V\rangle_3)$. Both projections are implemented by means of quarter-wave plates, half-wave plates and polarizing cubes. The photons are then filtered by \SI{10}{\nano\meter} interference filters, coupled into single--mode optical fibers and brought to single-photon detectors. Motorized translations M$_{1}$ and M$_{2}$ ensure temporal overlap of the photons on PBS and FBS, respectively.

\section{Measurement and results}
We have measured the collectibility on three characteristic quantum states: (i) $|\psi_{1}\rangle = \tfrac{1}{\sqrt{2}}(|00\rangle + |11\rangle)$ (Bell state), (ii) $ |\psi_{2}\rangle = |10\rangle$ (pure separable state), (iii) $\hat{\rho}_{3} = \tfrac{1}{4}\hat{\mathds{1}}$ (maximally mixed state), which were encoded twice in the following HESs
\begin{equation}
\begin{split}
|\Psi_{x}\rangle_{\mathrm{HES}}  & = |\psi_x\rangle_{Ap,Bp} \otimes |\psi_x\rangle_{As,Bs}\quad\mbox{for $x=1,2$,}\\
\hat{\rho}_{3,\mathrm{HES}} & = \tfrac{1}{16}\,\hat{\mathds{1}}_{Ap,Bp,As,Bs}.
\end{split}
\label{genstates}
\end{equation}
States (\ref{genstates}) generated in the experiment can be expressed via Eq.~(\ref{eq:encode}) as
\begin{equation}
\begin{split}
|\Psi_{1}\rangle_{\mathrm{HES}} & = \tfrac{1}{2}(|HH\rangle_{1,3} + |VH\rangle_{2,3}+|HV\rangle_{1,4} + |VV\rangle_{2,4})\\
|\Psi_{2}\rangle_{\mathrm{HES}} & = |VH\rangle_{2,3},\\
\hat{\rho}_{3,\mathrm{HES}} & = \tfrac{1}{16}\,\hat{\mathds{1}}^{(p)}\otimes \hat{\mathds{1}}^{(s)}, 
\end{split}
\label{states}
\end{equation}
which were prepared by suitable choice of pumping beam polarization, rotation of HWP$_{1,2}$ and for hyper--entangled Bell state ensuring the overlap on PBS.
 
\begin{figure}[t]
                \begin{center}
                \includegraphics[scale=1]{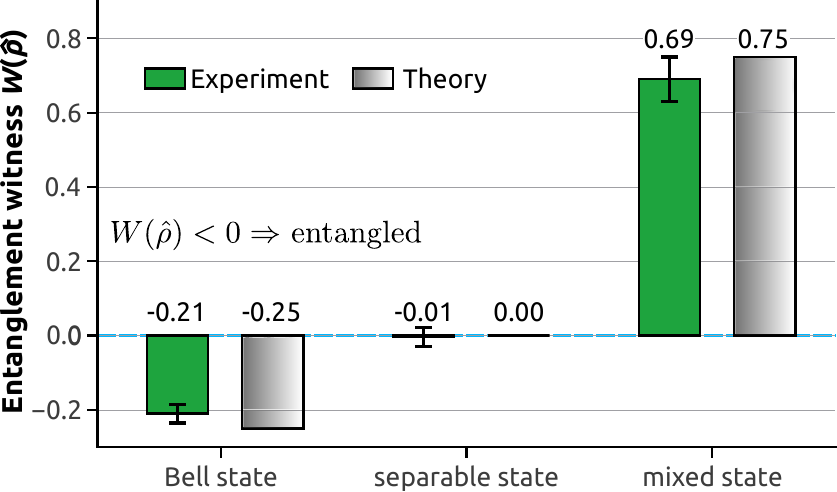}
                \caption{\label{hypercol}(color online) Experimental results and theoretical values of entanglement witness for three characteristic quantum states. Note that measurement values are in good agreement with the theoretically predicted values. As expected, the witness gives a clearly negative values only in the case of the Bell state.}    
                \end{center}
\end{figure}

For the Bell and mixed state, we were able to tune the probability $P$ to $P = 0.50,$ and for the separable state to $P = 0.01$. The values of $P$ are adjusted using purely single--photon detection events and as a result the uncertainties of $P$ are negligible in comparison with the uncertainty of two--photon coincidence detections. Obtained experimental and theoretically calculated values of collectibility $W(\hat{\rho})$ for the states from (\ref{states}) are summarized in Tab.~\ref{tab1} and visualized in Fig.~\ref{hypercol}.

\begin{table}
\caption{Summarized measured values of $W$ and their theoretical predictions $W_{\mathrm{th}}$ obtained for the states defined in Eq.~(\ref{states}).}
\begin{ruledtabular}
\begin{tabular}{ccc}
    Quantum state & $W$ & $W_{\mathrm{th}}$\\\hline
    Bell state ($|\Psi_{1}\rangle_{\mathrm{HES}}$)& -0.21 $\pm$ 0.03 & -0.25 \\
    Separable state ($|\Psi_{2}\rangle_{\mathrm{HES}}$) & -0.01 $\pm$ 0.03 & 0.00 \\
    Mixed state ($\hat{\rho}_{3,\mathrm{HES}}$) & 0.69 $\pm$ 0.06 & 0.75 \\
\end{tabular}
\label{tab1}
\end{ruledtabular}
\end{table}

\begin{figure}[t]
                \begin{center}
                \includegraphics[scale=1]{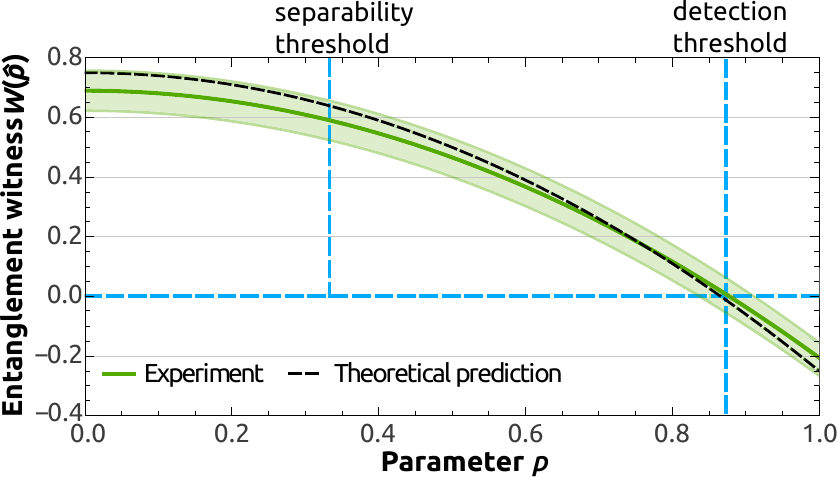}
                \caption{\label{werner} (color online) Experimental results and theoretical prediction of collectibility for the Werner states as a function of parameter $p$. The solid green line depicts the observed experimental results, the dashed black line is the theoretical prediction and the light green area corresponds to the measurement uncertainty.}    
                \end{center}
\end{figure}

Further, we have investigated the collectibility of Werner states which up to local unitary transformations can be expressed in form of a weighted sum of maximally entangled and maximally mixed state
\begin{equation}
\hat{\rho}_{W} = p|\psi_{1}\rangle\langle \psi_{1}| + q\hat{\rho}_{3},
\label{wstate}
\end{equation}
where $0\le p\le 1$ and $q=1-p.$ It follows from Eq.~(\ref{wstate}) that $p^{2}$ ($q^2$) is the probability of observing the maximally entangled (mixed) state simultaneously in both DOFs. The cross probability $2pq$ corresponds to observing entangled state in one DOF and mixed state in the other DOF. It can be directly shown that for a mixed state encoded into one DOF, the projection probabilities $p_{ij}$ and $p_{++}$ are independent of the state encoded into the other DOF. As a consequence, to obtain the values of collectibility for Werner states,
one can simply interpolate the results for $|\Psi_{1}\rangle$ and $\hat{\rho}_{3}$ with effective weights of $p^{2}$ and $1-p^{2},$ respectively~\cite{col1d}. Note that the collective entanglement witness is able to detect entanglement only for $p > \frac{\sqrt{3}}{2}$ although Werner states are already entangled for any $p > \frac{1}{3}$ \cite{Werner, Bovino,col1d,BartkiewiczPRA17a}. Obtained experimental and theoretical values of collectibility $W(\hat{\rho})$ as a function of the parameter $p$ are summarized in Tab.~\ref{tab2} and visualized in Fig.~\ref{werner}.

\begin{table}
\caption{Assorted results obtained for Werner states defined in Eq.~(\ref{wstate}), where $W$ is the observed value of entanglement witness and $W_{\mathrm{th}}$ is its theoretical prediction.}
\begin{ruledtabular}
\begin{tabular}{ccc}
    $p$ & $W$ & $W_{\mathrm{th}}$\\
    \hline
    0.0 & 0.69 $\pm$ 0.06 & 0.75 \\
    0.2 & 0.65 $\pm$ 0.06 & 0.71 \\
    0.4 & 0.55 $\pm$ 0.06 & 0.59 \\
    0.6 & 0.37 $\pm$ 0.06 & 0.39 \\
    0.8 & 0.11 $\pm$ 0.06 & 0.11 \\
    1.0 & -0.21 $\pm$ 0.03 & -0.25 \\
\end{tabular}
\label{tab2}
\end{ruledtabular}
\end{table}

\section{Quick quality check of HES}
\begin{figure}[t]
                \begin{center}
                \includegraphics[scale=1]{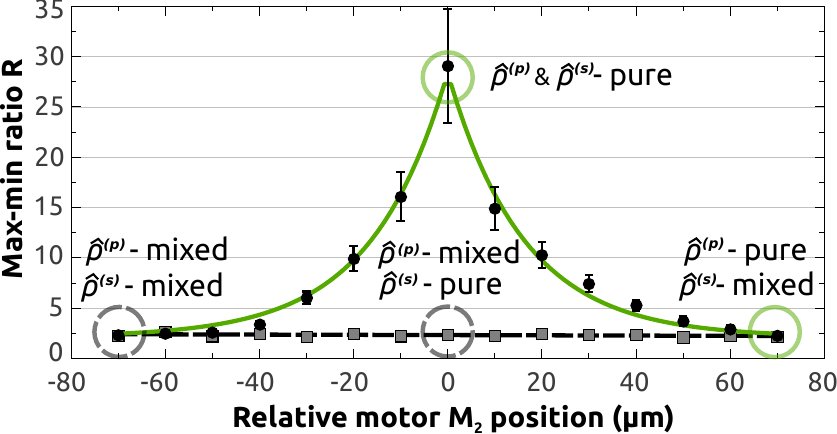}
                \caption{\label{ratio} (color online) Experimental values of the ratio $R$ [see Eq.~(\ref{eq:R})] obtained for the quality analysis of HESs for Werner states. The solid green (dashed black) curve corresponds to pure (mixed) $\hat{\rho}^{(p)}$. The collectibility has been measured for the encircled states.}        
                \end{center}
\end{figure}

Here, we demonstrate that collectibility can be used to quickly check the quality of hyper entanglement. HES transmission through a noisy channel can result in decreased purity independently in both DOFs (i.e., in general $\hat{\rho}^{(p)}$ and $\hat{\rho}^{(s)}$ are  different). This effect non trivially affects the collectibility measurement. We obtained $\hat{\rho}^{(p)}$ and $\hat{\rho}^{(s)}$ states of different purities experimentally by intentionally detuning temporal overlap between the photons using motorized translations $M_{1}$ and $M_{2}$. The purity of the tested states does not affect the values of $p_{ij}$ [see Eq.~(\ref{projections})] that corresponds to local projections onto states $|H\rangle_{1(2)}, |V\rangle_{2}$ [see Eq.~(\ref{eq:encode})]. The only difference is observed for $p_{++}$ which is measured by local projection $\tfrac{1}{2}(|H\rangle_{1}+|V\rangle_{1}+|H\rangle_{2}+|V\rangle_{2})$.
Due to interference, this measurement is sensitive to phase difference between spatial modes $1$ and $2$. We define the ratio $R$ as function of this phase shift
\begin{equation}
R = \frac{cc_{max}}{cc_{min}},\label{eq:R}
\end{equation}
where $cc_{max}$ and $cc_{min}$ stand for maximum and minimum coincidences rates. The measurement was implemented by combining pure and mixed states $\hat{\rho}^{(p)}$  with pure and mixed states $\hat{\rho}^{(s)}$. The observed experimental values are summarized in Tab.~\ref{tab3} and visualized in Fig.~\ref{ratio}. Note that if a HES becomes disentangled in one DOF, the ratio $R$ goes to 1 (as seen in Fig.~\ref{ratio}). The value of $p_{++}$ is then $\frac{1}{4}$ and $W(\hat{\rho})$ is positive. On the other hand, when both $\hat{\rho}^{(p)}$ and $\hat{\rho}^{(s)}$ are sufficiently pure and entangled, $W(\hat{\rho})$ becomes negative. Hence, this method is a quick and easy way to diagnose HES distribution.

\begin{table}
\caption{The measured value of entanglement witness $W$ and its predicted theoretical value $W_{\mathrm{th}}$ for the quality analysis of HESs for Werner states.}
\begin{ruledtabular}
\begin{tabular}{cccc}
    $\rho^{(p)}_{W}$& $\rho^{(s)}_{W}$ & $W$ & $W_{\mathrm{th}}$\\[2mm]
    \hline
    pure & pure  & -0.21 $\pm$ 0.03 &  -0.25   \\
    pure & mixed  & 0.71 $\pm$ 0.06 &   0.75  \\
    mixed & pure  & 0.70 $\pm$ 0.06 &   0.75  \\
    mixed & mixed  & 0.69 $\pm$ 0.06 &   0.75 \\
\end{tabular}
\label{tab3}
\end{ruledtabular}
\end{table}

\section{Conclusions}
We have reported on experimental measurement of collective nonlinear entanglement witness known as the collectibility on a single copy of a HES. The obtained results are in good agreement with theoretical predictions. The collectibility witness for hyper--entangled Bell state ($-0.21 \pm 0.03$) is negative with sufficient certainty and also close to its theoretical value. As expected, the observed results of $W(\hat{\rho})$ for the separable and the mixed state are non--negative and within one standard deviation from theoretically calculated values. We have interpolated the collectibility witness for several Werner states. These experimental results conform with theoretically predicted connection between collectibility witness and the Werner states parameter $p$. Further, we have demonstrated the impact of states purities on the collectibility witness. The results show that if HES becomes disentangled in one or more degrees of freedom, the collectibility witness becomes positive. The method for diagnostics of HESs developed in this paper is not as robust as quantum state tomography but can be appealing for applications that need fast verification whether the quantum system is sufficiently hyper--entangled.
The experimental accessibility of our method makes it suitable for further development of other nonlinear EWs requiring more than two copies of the measured state \cite{BartkiewiczPRA15a,BartkiewiczPRA15b,BartkiewiczPRA17b} or it can be easily adapted to measure a class of two-copy based EWs studied in Ref.~\cite{BartkiewiczPRA17a}.

\begin{acknowledgments}
Authors thank Cesnet for providing data management services. Authors acknowledge
financial support by the Czech Science Foundation under the project No. 17-10003S. KB also acknowledges the financial support of the Polish National Science Center under grant No. DEC-2015/19/B/ST2/01999. The authors also acknowledge the projects Nos. LO1305 and CZ.02.1.01./0.0/0.0/16\textunderscore 019/0000754 of the Ministry of Education, Youth and Sports of the Czech Republic financing the infrastructure of their workplace and VT also acknowledges the Palacky University internal grant No. IGA-PrF-2018-009
\end{acknowledgments}

\end{document}